\documentstyle[twocolumn,aps]{revtex}             
\input epsf

\begin{document}
\title{New tests for a singularity of ideal MHD}
\author{Robert M. Kerr$^1$ and Axel Brandenburg$^2$}
\address{$^1$NCAR, Boulder, CO 80307-3000;
$^2$Mathematics, University of Newcastle, NE1 7RU, UK}

\maketitle

\begin{abstract}
Analysis using new calculations with 3 times the resolution of
the earlier linked magnetic flux tubes confirms the transition 
from singular to saturated
growth rate reported by Grauer and Marliani \cite{GrauerMar99} 
for the incompressible cases is confirmed. However, all of the
secondary tests point to a transition back to stronger growth rate
at a different location at late times.  
Similar problems in ideal hydrodynamics are discussed, pointing out
that initial negative results eventually led to better
initial conditions that did show evidence for a singularity of Euler.
Whether singular or near-singular growth in ideal MHD is eventually
shown, this study could have bearing on
fast magnetic reconnection, high energy
particle production and coronal heating.
\end{abstract}
The issue currently leading to conflicting conclusions about
ideal 3D, incompressible MHD is similar \cite{KerrB99,GrauerMar99}
to what led to conflicting results
on whether there is a singularity of the 3D incompressible Euler.
With numerical simulations, it was first concluded that 
uniform mesh calculations with symmetric initial conditions such
as 3D Taylor-Green were not yet singular \cite{Brachetetal83}.  
Next, a preliminary 
spectral calculation \cite{KerrH89} found weak evidence in
favor a singularity in a series of Navier-Stokes simulations
at increasing Reynolds numbers, but larger
adaptive mesh or refined mesh calculations 
did not support this result \cite{PumirS90,Shelleyetal93}. 
Eventually, numerical evidence in favor of
a singularity of Euler was obtained using several independent tests applied
to highly resolved, refined mesh
calculations of the evolution of two anti-parallel vortex tubes \cite{Kerr93}.
To date, these calculations have met every analytic test for whether
there could be a singularity of Euler.

Several other calculations have also claimed numerical
evidence for a singularity of Euler \cite{BoratavPZ92,BoratavP94,GrauerMG98}.
While in all of these cases the evidence is plausible, with the perturbed
cylindrical shear flow \cite{GrauerMG98} using the BKM
$\|\omega\|_\infty$ test \cite{BKM84}, for none has
the entire battery of tests used for the anti-parallel case been applied.
We have recently repeated one of the orthogonal cases \cite{BoratavPZ92}
and have applied the BKM test successfully.  In all cases using the BKM
test, $|\omega\|_\infty\approx A/(T_c-t)$ with $A\approx19$.

To be able to make a convincing case for the existence of a singularity
in higher dimensional partial differential equations, great care must be
taken with initial conditions, demonstrating numerical convergence, and
comparisons to all known analytic or empirical tests.  On the other hand,
if no singularity is suspected, some quantity that clearly saturates should be
demonstrated, such as the strain causing vorticity growth \cite{PumirS90}. 
It is an even more delicate matter to claim that someone else's calculations or
conclusions are incorrect.  If it is a matter of suspecting there is
inadequate resolution, one must attempt to reproduce the suspicious 
calculations as nearly as possible
and show where inadequate resolution begins to
corrupt the calculations and how improved resolution changes the results.

\begin{figure}[t]
\epsfxsize=8.5cm
\epsfbox{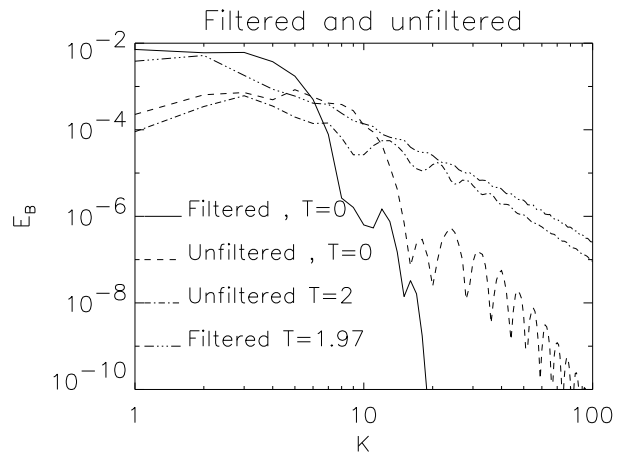}
\caption[]{Filtered and unfiltered initial and final spectrum .
The unfiltered spectrum is initialized on a $384^3$ mesh.}
\label{fig:spectra}\end{figure}

An example of how a detailed search for numerical errors
should be conducted can be found
in the extensive conference proceeding \cite{Kerr92} that appeared 
prior to the publication of the major results supporting the existence
of a singularity of Euler for anti-parallel vortex tubes \cite{Kerr93}.
The primary difference with earlier work 
was in the initial conditions.  It was found that 
compact profiles \cite{MelanderH89} were an improvement, 
but only if used in conjunction with a high wavenumber filter.  
Otherwise, the initial unfiltered energy spectrum of 
the bent anti-parallel vortex tubes went as $k^{-2}$. 
Oscillations in the spectrum at high wavenumber
in unfiltered initial conditions for linked magnetic flux tubes are shown
in Figure~\ref{fig:spectra}, showing that the initial MHD spectrum is steep
enough that eventually these oscillations are not important.

The purpose of this letter is to address the claim that a new adaptive
mesh refinement (AMR) calculation by Grauer and Marliani \cite{GrauerMar99}
supercedes our uniform mesh calculations \cite{KerrB99}
and that eventually there is a transition to exponential growth.
Note that this claim was made without any evidence for whether
their numerical method was converged.  In all of our earlier calculations,
once the calculations become underresolved, we also saw transitions
to exponential growth.

Not knowing exactly the initial condition used 
by the new AMR calculations \cite{GrauerMar99},
where and how much grid refinement was used, 
and the short notice we have been given to reply has proven a challenge.
Fortunately, we were in the process of new $648^3$ calculations in
a smaller domain of $4.3^3$, yielding effectively 3 times the local 
resolution of our earlier work \cite{KerrB99} in a $(2\pi)^3$ domain
on a $384^3$ mesh.  The case with an initial flux tube diameter of
$d=0.65$, so that the tubes slightly overlap, 
appears to be closer to their initial condition
and so will be the focus of this letter.
The importance of our other initial condition, with $d=0.5$, and
no initial overlap of the tubes, is that it is less influenced by 
an initial current sheet that forms near the origin and is claimed
to be the source of the saturation of the nonlinear terms.  This
was used for the compressible calculations.

\begin{figure}[t]
\epsfxsize=8.5cm
\epsfbox{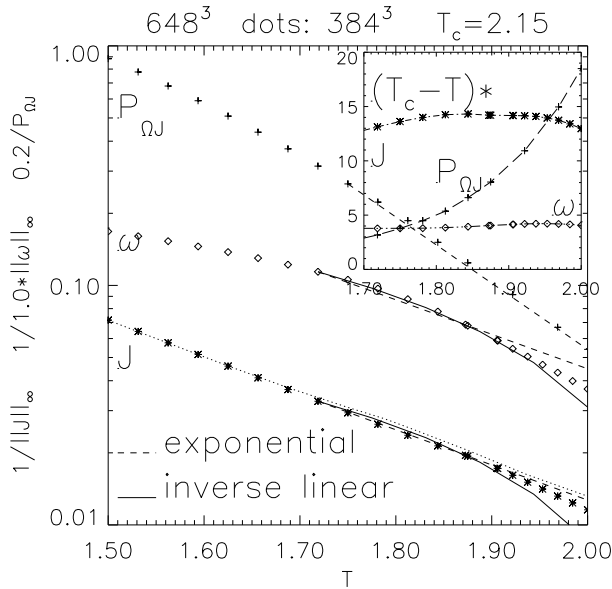}
\caption[]{Replot of $\|J\|_\infty$, $\|\omega\|_\infty$,
and $P_{\Omega J}$ for the new incompressible calculations
on $4.3^3$ domain with initial condition $d=0.65$
in semi-log coordinates.  All plots are from the $648^3$ calculation
except one $384^3$ plot of $1/\|J\|$.  Exponential and inverse linear
fits are shown for $t=1.75$ to 1.98.  
Each works equally well for $\|J\|_\infty$,
inverse linear is better for $\|\omega\|_\infty$, and exponential
is better for $P_{\Omega J}$.  Multiplying $\|J\|_\infty$,
$\|\omega\|_\infty$, and $P_{\Omega J}$ by $(T_c-t)$ in the
inset emphasizes that $\|J\|_\infty$ and $\|\omega\|_\infty$ might
be showing consistent singular behavior.  The large growth of $P_{\Omega J}$
is discussed.}
\label{fig:incomexp}\end{figure}

\begin{figure}[t]
\epsfxsize=8.5cm
\epsfbox{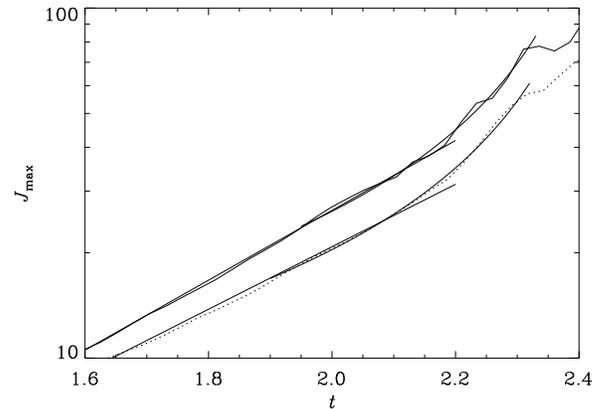}
\caption[]
{Semi-logarithmic plot of $\|J\|_\infty$ for a compressible $240^3$
calculation in a domain of size 4 (dotted line: filtered, and solid
line: unfiltered initial conditions) together with fits to exponential
growth and blow-up behavior, respectively. The latter are better fits
at later times.}
\label{fig:comprexp}\end{figure}

Using semi-log coordinates, Figure \ref{fig:incomexp} plots the growth of
$\|\omega\|_\infty$ and $\|J\|_\infty$ for our new high resolution
incompressible calculation and Figure \ref{fig:comprexp} plots
$\|J\|_\infty$ for a new compressible calculation.
By taking the last time all relevant quantities on
the $384^3$ and $648^3$ grids were converged, $\|J\|$ being the worst, 
then by assuming that the smallest scales are decreasing linearly towards a
possible singular time, an estimate of the last time the $648^3$ calculation
was valid was made.  To test exponential versus
inverse linear growth, fits were taken between $T=1.72$ and 1.87, then
extrapolated to large $T$.  The large figure shows
that either an exponential or a singular $1/(T_c-t)$ form could fit the data, 
while the inset shows that taking an estimated singular time of 
$T_c=2.15$ and multiplying by $(T_c-T)$ that at least
$\|J\|_\infty$ and $\|\omega\|_\infty$ have consistent singular behavior
over this time span.
The strong growth of $P_{\Omega J}=\int \,dV(\omega_i e_{ij} \omega_j
- \omega_i d_{ij} J_j - 2\varepsilon_{ijk}J_i d_{j{\ell}}e_{{\ell}k})$,
which is the production of $\int dV (\omega^2+J^2)$, is discussed below.  
The $384^3$ curve for $1/\|J\|$ demonstrates that lack of resolution
tends to exaggerate exponential growth.  For the compressible
calculations it can be seen that there also is an exponential regime
that changes into a regime with $1/\|J\|_\infty\sim (T_c-t)$.

Using the new incompressible calculations and applying the 
entire battery of tests, based upon Figure \ref{fig:incomexp}
we would agree that for the incompressible case there is
a transition as reported \cite{GrauerMar99} and signs of
saturation at this stage are shown below.  
Whether the transition is to exponential for all times
as claimed \cite{GrauerMar99}, 
or whether there is a still later transition to different singular behavior,
will be the focus of this letter.  We will look more closely at the
structure of the  current sheet we all agree exists \cite{KerrB99,GrauerMar99}
for signs of saturation.

The case against a singularity in
early calculations of Euler \cite{PumirS90,PumirKerr87,Brachetetal92} was
the appearance of vortex sheets, and through analogies with 
the current in 2D ideal MHD, a suggestion that this leads to a depletion
of nonlinearity.  The fluid flow most relevant to the
linked flux rings is 3D Taylor-Green, due to the initial symmetries
\cite{Brachetetal83}.  For both TG and linked flux tubes, 
two sets of anti-parallel vortex pairs form that are skewed with
respect to each other and are colliding.  In TG, just after the
anti-parallel vortex tubes form there is a period of slightly
singular development.  This is suppressed once the pairs
collide with each other, and then vortex sheets dominate for
a period.  The vortex sheets are very thin, but go across the
domain, so fine localized resolution might not be an advantage at this stage.
At late phases in TG,
the ends of the colliding pairs begin to interact with each other,
so that at 4 corners locally orthogonal vortices begin to form.
Due to resolution limitations, an Euler calculation of
Taylor-Green has not been continued long enough to determine whether,
during this phase, singular behavior might develop.  
We would draw a similar conclusion for all of MHD cases studied to date
\cite{GrauerMar99,Politanoetal95,Onoetal96}, that
there might not be enough local resolution to draw any final conclusions
even if AMR is applied.

While Taylor-Green has not been continued far enough to rule out
singularities, the final
arrangement of vortex structures led first to studies of
interacting orthogonal vortices \cite{BoratavPZ92}, 
and then anti-parallel vortices (see references in \cite{Kerr93}).
Both of these initial conditions now appear to develop singular behavior.
An important piece of evidence for a singularity of Euler
was that near the point of a possible singularity, the structure could
not be described simply as a vortex sheet.  Therefore,
there is a precedent to earlier work suggesting sheets, suppression
of nonlinearity, and no singularities to later work showing fully
three-dimensional structure and singular behavior. 

The initial singular growth of $\|J\|_\infty$ and 
$\|\omega\|_\infty$ for the linked flux rings,
then the transition to a saturated growth rate, might be due to
the same skewed, anti-parallel vortex pair interaction
as in Taylor-Green.  Even if this is all that is happening, the
strong initial vorticity production and shorter dynamical timescale
(order of a few Alfv\'en times)
than earlier magnetic reconnection simulations with anti-parallel
flux tubes \cite{Onoetal96} is a significant success of these simulations.
It might be that the vortices that have been generated are strong enough
to develop their own Euler singularity.  However, the interesting
physics is how the magnetic field and current interact with the
vorticity.  Do they suppress the tendency of the vorticity to become
singular, or augment that tendency?

\begin{figure}[t]
\epsfxsize=8.5cm
\epsfbox{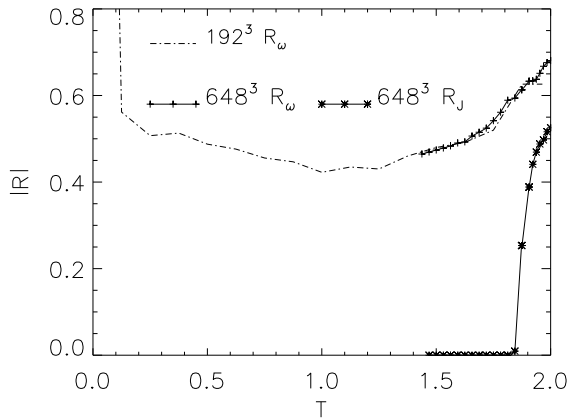}
\caption[]{Positions of $\|J\|_\infty$ and $\|\omega\|_\infty$ for
$d=0.65$ in a $4.3^3$ domain.} 
\label{fig:pos65}\end{figure}
One sign for saturation of the linked flux ring interaction would be
if the strongest current remains at the origin in this sheet.
Figure \ref{fig:pos65} plots the positions of $\|J\|_\infty$ 
and $\|\omega\|_\infty$
from the origin as a function of time.  During the period where 
exponential growth is claimed \cite{GrauerMar99}, $\|J\|_\infty$
is at the origin, which would support the claims of saturation.  However,
this situation does not persist.

By analogy to the movement of the $L^\infty$ norms of the components
of the stress tensor $u_{i,j}$ in Euler, we expect that the positions of 
$\|J\|_\infty$ and $\|\omega\|_\infty$
should approach each other and an extrapolated singular point 
in ideal MHD. 
Figure \ref{fig:pos65} supports the prediction that the positions of
$\|J\|_\infty$ and $\|\omega\|_\infty$ should approach each other
but so far not in a convincingly linear fashion.  This is addressed next.
We have similar trends for the positions of $\|J\|_\infty$ and
$\|\omega\|_\infty$ in the compressible calculations.

\begin{figure}[t]
\epsfxsize=8.5cm
\epsfbox{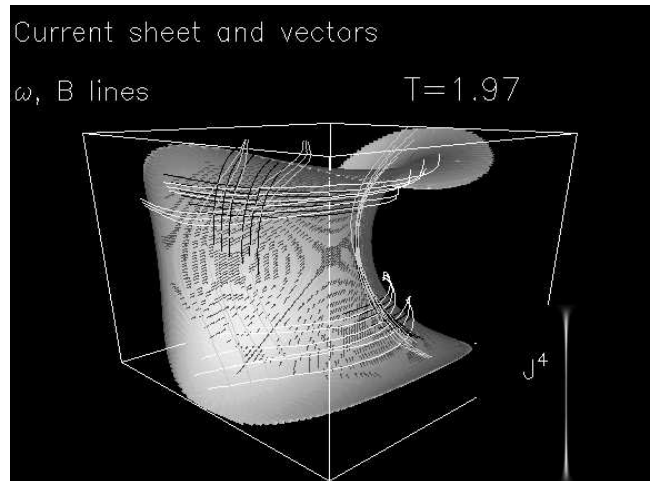}
\caption[]{For $t=1.97$ on the inner $162^3$ grid points, the current sheet
is shown with arrows of $\overrightarrow{J}$ overlaid in dark. 
The current through the $(x/y=z)$ plane containing $\|J\|_\infty$ is in 
lower right. Contours of $|J|^4$ are shown to emphasize where $\|J\|_\infty$
is located.  Dark lines are 
$\overrightarrow{B}$ and light lines are $\overrightarrow{\omega}$ 
that originated in the vicinity of $\|J\|_\infty$.
The vortex lines are predominantly those in the double vortex
rings that were originally generated by the Lorenz force, then
became responsible for spreading out the current sheet. 
Where the $\overrightarrow{B}$ lines cross in the upper left and
lower right corners are around the locations of $\|J\|_\infty$, which
due to symmetries are different views of the same structure.  
Near $\|J\|_\infty$, $\overrightarrow{B}$ nearly overlies and is parallel to
$\overrightarrow{\omega}$ and both
$\overrightarrow{B}$ and $\overrightarrow{\omega}$ are nearly orthogonal to
their partners across the current sheet, 
where $\overrightarrow{B}$ and $\overrightarrow{\omega}$
are anti-parallel.  Taken from the $d=0.65$ calculation in a $4.3^3$ domain
on a $648^3$ mesh.}
\label{fig:Job}\end{figure}

Figure \ref{fig:Job} gives an overall view of the current, vorticity 
and magnetic field around the inner current sheet.  The vortex pattern has
developed out of the four initial vortices, two sets of orthogonal, 
anti-parallel pairs that are responsible for the initial compression and 
stretching of the current sheet.  By this late time, the ends of the those
vortices have begun to interact as new sets of orthogonal vortex pairs.
The lower right inset in Figure~\ref{fig:Job} is a 2D $x/(y=z)$ slice
through this domain that goes through $\|J\|_\infty$ at
$t=1.97$ to show that while $\|J\|_\infty$ is large at the origin $(0,0,0)$, 
$\|J\|_\infty$ is larger where it 
is being squeezed between the new orthogonal vortices.
Along one of the new vortices $\overrightarrow{B}$ is parallel to and overlying
$\overrightarrow{\omega}$ and on the orthogonal partner they are
anti-parallel and overlying.

The location of $\|\omega\|_\infty$ is not in the vortex lines
shown, but is on the outer edges of the current sheet.
Therefore, the exact position of $\|\omega\|_\infty$ in Figure 
\ref{fig:pos65} is an artifact of the initial development and does not 
accurately reflect the position of $\overrightarrow{\omega}$ most directly
involved in amplifying $\|J\|_\infty$, which is probably why
the positions of $\|J\|_\infty$ and $\|\omega\|_\infty$ are not
approaching each other faster.  The continuing effects of the initial
current sheet is probably also behind the strong exponential growth
of $P_{\Omega J}$ in Figure~\ref{fig:incomexp}, stronger even than the
the possible singular growth of $\|J\|_\infty$ and
$\|\omega\|_\infty$ in the inset.  More detailed
analysis in progress should show that near the position of $\|J\|_\infty$,
the growth of $P_{\Omega J}$ and the position of $\|\omega\|_\infty$ are
more consistent with our expectations for singular growth and has
already shown that some of the components of $P_{\Omega J}$ have consistent
singular growth.

As noted, for Euler all available calculations find
$|\omega\|_\infty\approx A/(T_c-t)$ with $A\approx19$.  $A$ represents
how much smaller the strain along $\|\omega\|_\infty$ is than
$\|\omega\|_\infty$. Here, $A\approx 4$, indicating stronger growth
in $\|\omega\|_\infty$ for ideal MHD than Euler.  
Another Euler result was that the asymptotic energy spectrum as the
possible singularity was approached was $k^{-3}$, whereas purely
sheet-like structures in vorticity should yield $k^{-4}$ spectrum.
$k^{-3}$ indicates a more complicated 3D structure than sheets.
In Figure~\ref{fig:spectra} the late time spectra are again $k^{-3}$.

The next the initial condition we will investigate will be
magnetic flux and vortex tubes that
nearly overlay each other and are orthogonal to their partners.
Our new calculations of orthogonal vortex tubes for Euler show
that they start becoming singular as filaments are pulled off
of the original tubes and these filaments become anti-parallel,
suggesting that the fundamental singular interaction in Euler
is between anti-parallel vortices.  Whether the next step for
ideal MHD is to become anti-parallel or something else can only
be determined by new calculations.
AMR might be useful, but great care must be taken with the placement
of the inner domains and a large mesh will still be necessary.
The complicated structures in the domain in Figure \ref{fig:Job}
are not fully contained in this innermost $162^3$ mesh points and
the innermost domain should go out to the order of $300^3$ points.
There are examples of how to use AMR when there are
strong shears on the boundaries of sharp structures \cite{GrabowClark93}.
This uncertainty of where to place the mesh is why we believe in
using uniform mesh calculations as an unbiased first look at the problem.

These final results are hardly robust and their usefulness
is primarily to suggest a new more localized initial condition and to show
that none of the calculations to date is the full story. 
For $J$ and $\omega$ to show singular behavior
as long as they have has been surprising.  
Recall that for Euler, velocity, vorticity and strain are all 
manifestations of the same vector field, but for ideal MHD there
are two independent vector fields even though the only analytic
result in 3D is a condition on the combination, 
$\int dV \left[\|\mbox{\boldmath$\omega$}\|_\infty(t)+\|{\bf J}\|_\infty(t)
\right] dt \rightarrow\infty$ \cite{CKS97}.  
Eventually, one piece of evidence
for singular growth must be a demonstration of strong coupling between
the current and vorticity so that they are acting as one vector field.
It could be that our strong growth is due to the strongly helical 
initial conditions and there are no singularities.  This would still
be physically interesting since helical conditions 
could be set up by footpoint motion in the corona.

Could the magnetic and electric fields blow up too?
There are signs this might be developing around the final position of
$\|J\|_\infty$, in which case there might exist a mechanism
for the direct acceleration of high energy particles.  
This has been considered on larger scale \cite{Milleretal97}, but
to our knowledge a mechanism for small-scale production of super-Dreicer
electric fields has not been proposed before.  A singular rise
in electric fields could explain the sharp rise times in X-ray
production in solar coronal measurements \cite{Kiplingeretal84}, 
which could be a consequence 
of particle acceleration coming from reconnection.  This
would also have implications for the heating of the
solar corona by nanoflares \cite{Parker81} and the production of cosmic rays.

This work has been supported in part by an EPSRC visiting grant GR/M46136.
NCAR is support by the National Science Foundation.

\end{document}